\documentclass[aps,prb,twocolumn,groupedaddress,showpacs,superscriptaddress,amssymb,amsmath]{revtex4-1}
\usepackage{graphicx}
\usepackage{amsmath}
\usepackage{amssymb}
\usepackage{amsfonts}
\usepackage{color}
\usepackage{bm}        
\usepackage{comment}
\usepackage{placeins}
\usepackage[mathscr]{eucal}
\usepackage[colorlinks, linkcolor=red,citecolor=blue,urlcolor=blue]{hyperref}
\usepackage[normalem]{ulem}
\usepackage[all]{hypcap}
\usepackage{multirow}
\usepackage[dvipsnames,usenames,table]{xcolor}
\usepackage{dcolumn}
\usepackage{hyperref}
\usepackage{bm}
\usepackage{epsf}
\usepackage{subfigure}
\usepackage{amsfonts}
\usepackage{epstopdf}%
\usepackage{braket}
\usepackage{float}
\newcommand{\be}{\begin{equation}}
\newcommand{\ee}{\end{equation}}
\newcommand{\bea}{\begin{eqnarray}}
\newcommand{\eea}{\end{eqnarray}}
\begin{document}
\title{Exact Results for Interacting Hard Rigid Rotors on a d-Dimensional Lattice}
\author{Sushant Saryal}
\email{sushant.saryal@students.iiserpune.ac.in}
\affiliation{Department of Physics,Indian Institute of Science Education and Research, Pune 411008, India}
		
\author{Deepak Dhar}
\email{deepak@iiserpune.ac.in}	
\affiliation{Department of Physics,Indian Institute of Science Education and Research, Pune 411008, India}
		
\date{\today}

\begin{abstract}
We study the entropy of a set of identical hard objects, of general shape,  with each object pivoted  on the vertices of a d-dimensional regular lattice of lattice spacing $a$, but can have arbitrary orientations. When the pivoting point is situated asymmetrically
on the object, we show that there is a range of lattice  spacings $a$, where  in any orientation, a particle can overlap with at most one of its neighbors.  In this range, the entropy of the system of particles can be expressed exactly in terms of the grand partition function of  coverings of  the base lattice   by dimers at a finite negative activity. The   exact  entropy  in this range  is fully determined by the second virial coefficient.  Calculation of the partition function is also shown to be reducible to that of  the same model with discretized orientations.  We determine the exact functional form of the probability  distribution function of orientations at a site.  This depends on the density of  dimers  for the given activity in the dimer problem, that we determine by summing the corresponding Mayer series numerically.  These results are verified by numerical simulations. 
\end{abstract}

\maketitle

Many organic liquids, on cooling, first freeze into a state where there is long range crystalline order, but the molecules are partly able to rotate about their center of mass.  In contrast to the liquid crystals that shows orientational order, but no positional order, these materials show no long range orientational order, and  have been called plastic solids, rotator crystals or orientationally disordered solids.   Timmermans  studied it in the 1930's \cite{timmermans}, and early work has been summarized in  \cite{staveley}. 
These are currently  attracting  a lot of interest, for  applications such as storage batteries \cite{batteries},  drug delivery \cite{drug-delivery}, and    refrigeration \cite{ refrigeration}.

The simplest theoretical description of this state would be to assume that rigid molecules have  their centers of mass fixed at lattice positions, but have orientational degrees of freedom.  Many such models have been studied since the beginning of these studies, e.g.  \cite{pople-karasz}.  Casey and Runnels \cite{Runnels1} studied  a system of hard squares  with centers fixed on the one dimensional lattice and also  with centers free to move along a 1D line. Freasier and  Runnels\cite{Runnels2} studied a model of rotors with their centers pivoted on a one dimensional lattice. More recently, Kantor and  Kardar  \cite{Kardar}, and   Gurin  and Varga  \cite{Gurin} studied a system of hard rods in isobaric ensemble with their centers restricted to move only along a 1D line but free to rotate in a plane. \\
In an earlier paper, we have studied hard linear rods, with centers fixed on sites of a 1-d lattice, but free to rotate in a plane \cite{saryal1}.  We showed that even in one dimension, this system shows an infinite number of singularities of entropy per rod as a function of the ratio of length of rods and the lattice spacing. Plastic solids often show many (three or more) crystalline phases with varying amount of orientational order \cite{guthrie}.  It is   gratifying that the simple model of rigid hard particles can describe this basic phenomenology. 
\begin{figure}[h]
\centering
\includegraphics[width=0.6\columnwidth]{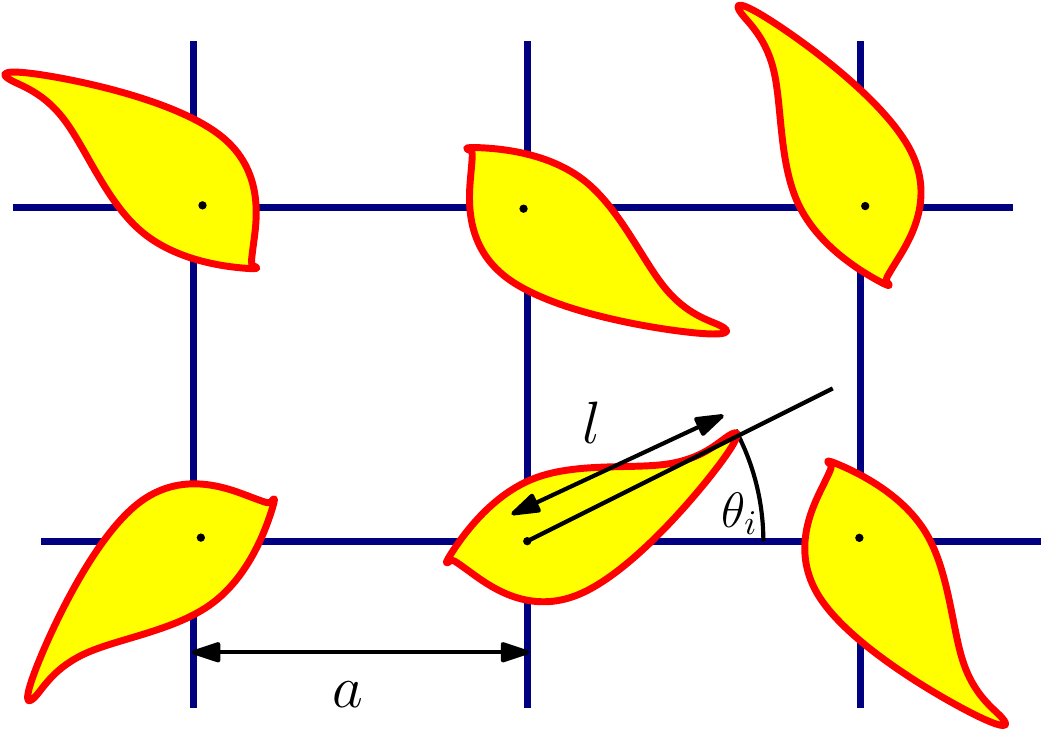}
\caption{An example of hard particles with their pivots fixed at sites of a square lattice of lattice spacing $a$. The farthest point in the object  from the pivot is at distance $\ell$. The orientation of the rotor at site ${\bf i}$ is specified by the orientation of the long axis to the x-axis $\theta_{\bf i}$.}
\label{fig1}
\end{figure}

\begin{figure}[h]
\centering
\includegraphics[width=0.8\columnwidth]{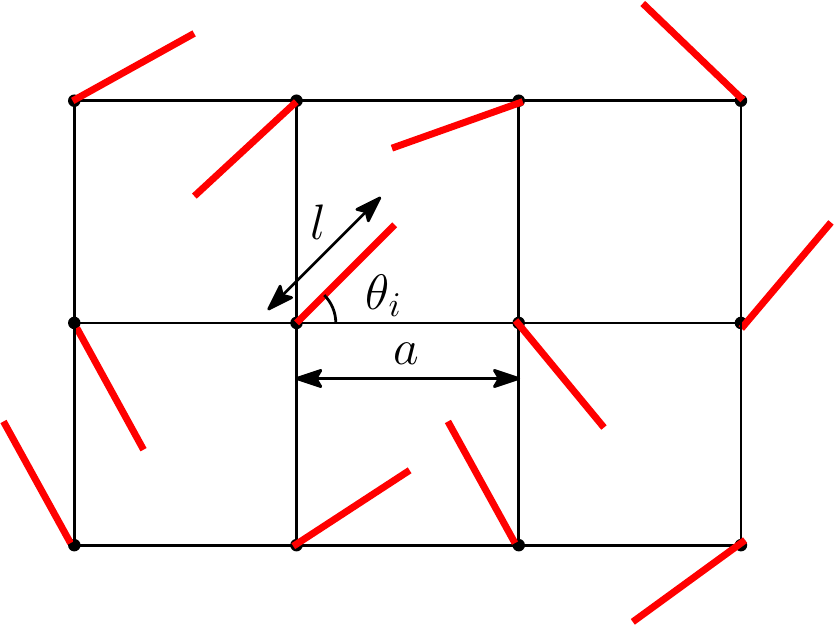}
\caption{Hard rods pivoted at the one end on a square lattice. Length of each rod is $l$ , lattice spacing is $a$. The orientation of rotor at site ${\bf i}$ is given by the angle its long axis makes with the x-axis.  }
\label{2drods}
\end{figure}

There are few  exact results available for this problem  in more than one dimension.  In this work, we show that in this model of rigid rotors on a lattice [an example in two dimensions is shown in Fig. \ref{fig1}], there is a range of lattice spacings in which the model becomes reducible to the dimer model on the same lattice with a negative value of the dimer activity. We determine the exact functional form of the probability  distribution function of orientations at a site.  This depends on the density of  dimers  for the given activity in the dimer problem, that we determine by summing the corresponding Mayer series numerically.  These results are valid for general shapes of rotors, and for arbitrary lattices, and dimensions. We have  verified these results by numerical simulations for thin  hard  linear rods pivoted at one end at the vertices of a square lattice [Fig. \ref{2drods}].\\
One important  feature of the model is the fact that interaction between two rotors has a different form in different directions. In some cases,  e.g. in the class of models known as compass models \cite{compass},  this feature  leads to the model becoming analytically tractable.  The best known example of this class is the Kitaev model \cite{kitaev}. In the  standard  version of the Kitaev model, the degrees of freedom are quantum spins, though the classical case has also been studied \cite{classical-kitaev}. We will only study the case of  classical rotors here.

We consider a system of  hard bodies, called rotors,    of identical shape and size in $d$ dimensions.  Each rotor has  a marked point called the pivot, which is at the same relative position in each rotor.  The objects are rigid, of a given shape.  Our treatment is valid for arbitrary shapes.  The pivots of the particles are fixed at the vertices  of a regular $d$-dimension lattice ( e.g.  the $d$-dimensional  hypercubical lattice), but the rotors can take arbitrary orientations,  subject to the constraint that different rotors cannot overlap.   The orientation of the rotor at the vertex $\bf{r}$ of the lattice may be specified by a $d \times d$  real orthogonal matrix $ M(\bf{r})$.  Then, the full configuration is specified by a lattice model, with a $d \times d$ matrix   belonging to the group $SO(d)$ at each vertex. [Sometimes, the experimentally prepared crystals are  racemic mixtures of  the two species of molecules related by mirror symmetry. These  crystals would be  described by the group O(d). Our treatment is easily extended to this case.]  

We first consider the case when the pivot is placed asymmetrically on the rotor, and there is a unique point on the rotor at maximum distance from the pivot. Let this distance be denoted by $\ell$.   Let the distance between nearest neighbors vertices of the lattice be $a$.The line joining the pivot to the farthest point will be called the long axis of the rotor. 
Then, we will study the entropy per site for this model, as a function of the dimensionless coupling parameter $\ell/a$, for a specified shape of the object.  

We define the indicator function $ \eta(\bf{a},\bf{b})$ to be $1$ if the rotors at sites $\bf{a}$ and $\bf{b}$, with given orientations $M(\bf{a})$ and $M(\bf{b})$, overlap, and zero otherwise. For notational simplicity, we will not  show the dependence on orientations explicitly. Then the partition function for this system is defined as
\begin{equation}
\mathcal{Z}_{N} = \left[ \prod_{\bf{r}} \int dM(\bf{r}) \right]  \prod_{<\bf{a} , \bf{b}>}\left[ 1 - \eta(\bf{a},\bf{b})\right],
\label{eq:Z}
\end{equation}
where the second product is over all nearest neighbor pairs, and the integral over the matrix $ M(\bf{r})$ is the normalized Haar measure over the group SO(d), so that $ \int dM({\bf r}) ~1 =1$.  We will only consider the case where rotors at sites farther than nearest neighbors cannot overlap, and second  product  is  restricted to only over nearest neighbor pairs.  The entropy per site $s(\ell/a)$ is defined by 
\begin{equation}
s(\ell/a) = \lim_{N \rightarrow \infty} [\log{Z}_N]/ N, 
\end{equation}
where $N$ is the number of sites in the lattice.
If $\ell/a < 1/2$, then different rotors cannot overlap, and the entropy per rotor takes its maximum value, zero  in  our normalization. As the spacing  $a$ is decreased, keeping $\ell$ constant, the entropy will also decrease. When $a$  is decreased to the minimum allowed value, corresponding to closest packing, entropy per rotor will tend to $-\infty$. 

 Now consider the case $ \ell/a = 1/2 + \epsilon$, where $\epsilon$ is a small positive number.  Consider two rotors at nearest neighbor vertices $\bf{i}$ and $\bf{j}$ along the bond  along one of the lattice directions, say  ${\bf \hat{e}}_1$.  Then,  these two rotors can overlap only if the long axes of both are substantially aligned parallel to the direction ${\bf \hat{e}}_1$. We define $\omega_1$ as set of orientations  of  the rotor at $\bf{i}$ such that 
there is a non-empty set of orientations of the rotor at $\bf{j}$ that overlap with the rotor at $\bf{i}$.
The volume of the set $\omega_1$ decreases to zero, as $\epsilon$ tends to zero. 

We similarly construct sets $\omega_{-1}, \omega_{\pm 2} , ...\omega_{\pm d}$, using other neighbors of $\bf{i}$ (set $\omega_{-\alpha}$ corresponds to the $-{\bf \hat{e}}_{\alpha}$ direction). The total number of such sets is the number of neighbors of $\bf{i}$, i.e. $2d$ on the $d$-dimensional hypercubical lattice. For small $\epsilon$, these sets $\omega_i$ are disjoint.  As $a$ is decreased, the sets $\omega_i$'s increase in size, and eventually, some will touch.  Let $a^*$ be the smallest value of $a$ such that the sets $\omega_i$ are still disjoint.  Then for all the range of values $2l > a > a^*$, a rotor at any site  $\bf{i}$, in any orientation $M(\bf{i})$
can have an overlap with at most one of its neighbors, whatever their orientations.  Let ${\bf j}$ and ${\bf j'}$ be two neighbors of ${\bf i}$, with ${\bf j} \neq {\bf j'}$. Then we have, for all ${\bf i},{\bf j},{\bf j}'$, and all  orientations $M({\bf i}), M({{\bf j}}), M(\bf{j}')$, 
\begin{equation}
\eta({\bf i}, {\bf j}) \eta({\bf i},{\bf j}') =0.
\end{equation}
We will call this the at-most-one-overlap ( AOO) condition.

A simple example of systems where AOO condition holds in some region of the parameter space  is hard rigid lines of length $\ell$, where one of the ends is the pivoting point as shown in Fig. \ref{2drods}. 
 In this case, it is straightforward to verify that  $ a^* = \sqrt{2} \ell$, and determine explicitly the function $\eta(\theta, \theta')$. The details are given in Appendix A.

\begin{figure}[H]
\centering
\includegraphics[width=0.6\columnwidth]{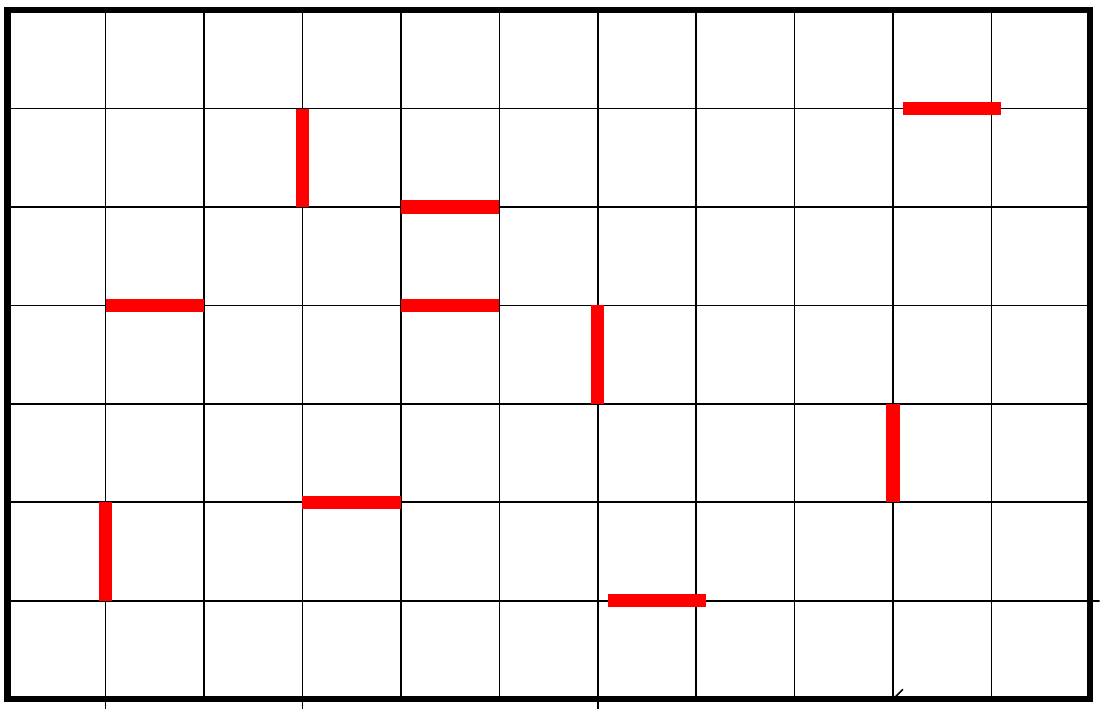}
\caption{A typical term in the graphical expansion of  the partition function [Eq.(\ref{eq:Z})]. This can also be thought of as a configuration  of a set of dimers on the lattice. }
\label{d1}
\end{figure}

\begin{figure}[H]
\centering
\includegraphics[width=0.8\columnwidth]{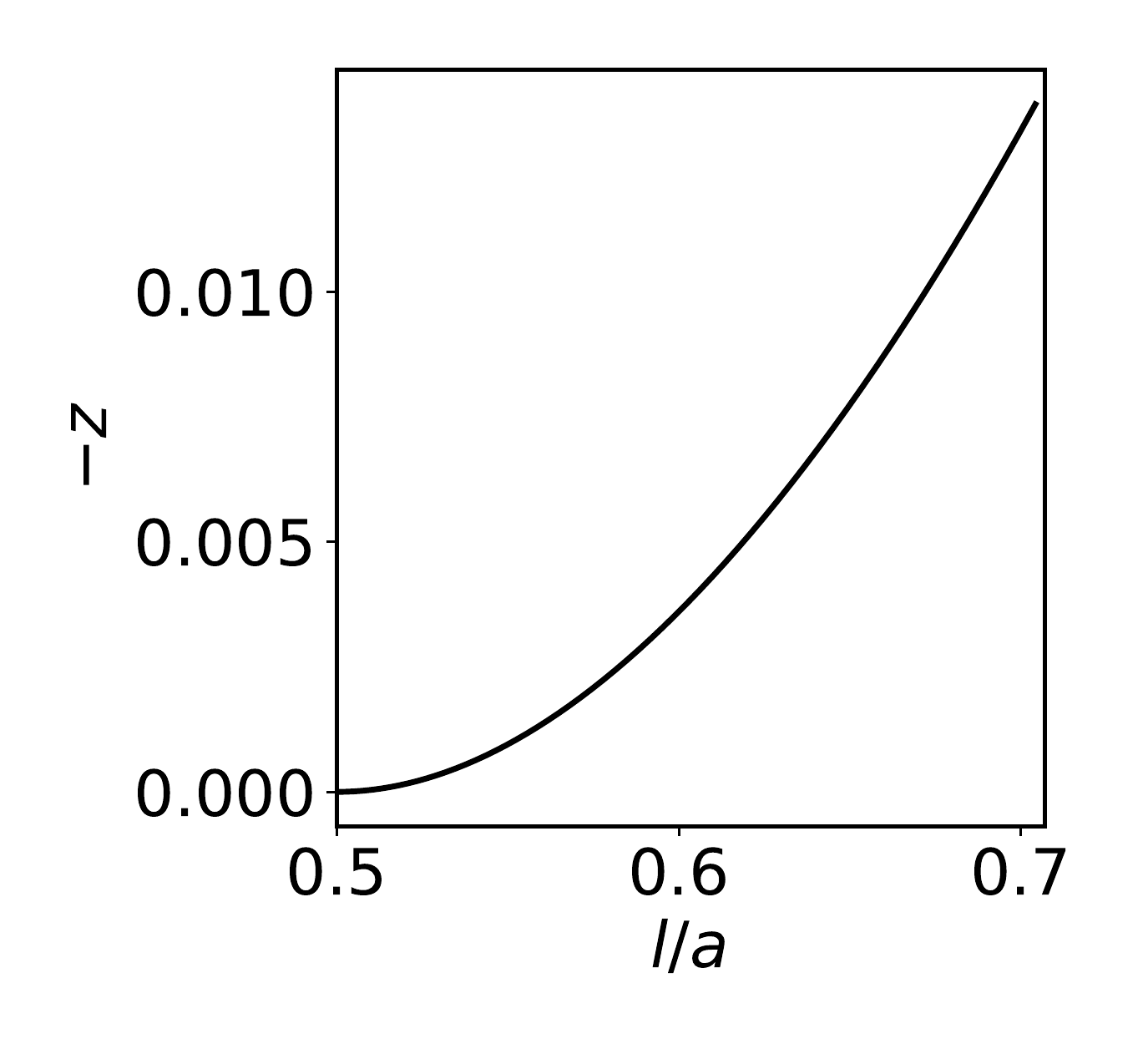}
\caption{$-z$ vs $l/a$, for a system of hard linear rods of length $l$ pivoted at one end on 2d square lattice. }
\label{za}
\end{figure}

Now, we expand the product in Eq. (\ref{eq:Z}), and make a graphical representation of the terms of the expansion, with a bond between vertices $\bf{i}$ and $\bf{j}$, iff the term contains $\eta(\bf{i},\bf{j})$, see Fig. \ref{d1}. This is a lattice version of the well-known  {\em Mayer cluster expansion}.  Then, using the AOO condition, only graphs  with at most one bond coming to a site survive.  Thus, we get
\begin{equation}
\mathcal{Z}_N = \mathcal{Z}_{dimers}(z),
\end{equation}
where $\mathcal{Z}_{dimers}(z)$ is the grand partition function of partial covering of the vertices of the lattice by dimers (see Fig. \ref{d1}), 
\begin{equation}
\mathcal{Z}_{dimers}(z)  =  \sum_{\textrm{dimer-confgs}} z^{\textrm{number of dimers}},
\end{equation}
with the fugacity of a dimer is the  (negative ) real number, given by 
\begin{equation}
z = - \int dM({\bf i}) \int dM({\bf j})  ~\eta({\bf i},{\bf j}).
\label{eq:z}
\end{equation}
In fact,  $z$ is exactly the second virial coefficient $B_2$ of the virial expansion. 
It is straight forward to determine it for a  a given shape of molecules. In Fig. \ref{za}, we have plotted  $z$ as a function of $\ell/a$ for the case of linear rods, pivoted at one end shown in Fig. \ref{2drods}. 
Let $g(z)$ be the logarithm  of the grand partition function per site, defined as 
\bea
g(z) = \lim_{N\rightarrow \infty}   [\log {\mathcal Z}_{dimers}(z)] / N
\eea
Then, the entropy per site is   $s(\ell/a) =g(z)$.

The evaluation of the partition function of partial covering of dimers is a well-known hard problem in lattice statistics, and no exact  expression for $g(z)$ is  known, except in one dimension and in  some  special  graphs, like the Bethe lattice. However, the equivalence is not without use, as we see that the exact entropy $s(\ell/a)$,  in the entire  range of $a$ satisfying the AOO condition,  depends on shape of the rotors, only through the second virial coefficient $z$.

Also, we note that the equality of  the dimer and rotor model partition functions holds also for finite lattices of size $L_1 \times L_2 \times L_3 ...$.  The dependence of these partition functions on one of the dimensions is of the form  
\begin{equation}
\mathcal {Z} \sim \sum_{\alpha} c_{\alpha} \lambda_{\alpha}^{L_1},
\end{equation}
where $\lambda_{\alpha}$ are the eigenvalues of the transfer matrix in the direction $1$, and $c_{\alpha}$ are some constants. Since this is true for all $L_1$, we conclude that {\em all} the nonzero eigenvalues of the transfer matrix for the rotor model are the same as eigenvalues of the transfer matrix for the dimer model with activity $z$. 
In Appendix B and C, we show an explicit calculation of the model for hard rods on a line and zigzag lattice respectively, using transfer matrices, and show that the results agree with the algebraic method used here.
We will now show that under the AOO condition,  our  model with continuous $SO(d)$ degrees of freedom is reducible to a discrete model with $2 d$ discrete degrees of freedom per site.   
 Following the terminology of ice and KDP models \cite{ice}, we will call this model the Hydrogen Iodide  (HI)  model, where the heavy ``iodine"(I) atoms sit at lattice sites, and there is one ``hydrogen"(H)  atom  near each I atom that sits along one of the nearest neighbor bonds, close to it.  There is a repulsive energy cost if the H atoms of two  adjacent I's sit on the same bond.  Thus, there is a discrete degree of freedom at each site ${\bf r}$ of the lattice, denoted by $\sigma(\bf r)$, which takes {\em non-zero} integer values between $-d$ and $d$. If the hydrogen atom is along the bond ${\bf \hat{e}}_{\alpha}$, we will call the state $\alpha$, if along $-{\bf \hat{e}}_{\alpha}$, we call the state $-\alpha$ ($1\leq \alpha \leq d$) .   
 Therefore,  for a bond between the sites ${\bf i}$ and ${\bf i}+ {\bf \hat{e}}_{\alpha}$, the energy cost is $J$ iff the spin at site ${\bf i}$ is in the state $\alpha$, and that at site $({\bf i} +{\bf \hat{e}}_{\alpha})$ is in the state $- \alpha$.   The interaction along the bond in the direction ${\bf \hat{e}}_{\alpha}$ only involves the spin states $\pm \alpha$.

The hamiltonian of the model is
\bea
{\mathcal H}_{HI} =  J \sum_{{\bf r}} \sum_{\alpha=1}^{d} \delta( \sigma({\bf r}), \alpha) \delta( \sigma({\bf r} + {\bf \hat{e}}_{\alpha}),  -\alpha), 
\eea
where $\delta(a,b)$ is the Kronecker delta function.   We write $ \exp \left[-\beta J \delta(a,b)  \right] = 1+ \tilde{z}  \delta(a,b)$, with $\tilde{z} = e^{ -\beta J} -1 $, \, $\beta$ is the inverse temperature, and expand the partition function in powers of $\tilde{z}$, we again get the partition functions of dimers with activity $z=\tilde{z}/(2d)^2$. 
Let $Z_{HI}$ denote the partition function of the HI model, then we have,
\bea
Z_{HI} = (2d)^N  {\mathcal Z}_{dimers} \Big( z = \frac{\tilde{z}}{(2d)^2}\Big)
\eea

 To complete the correspondence with the rotors models, we break the manifold of states given by $M$  into $2d$ equal  submanifolds.  A given matrix $M$ is associated with the discrete state $\alpha$ of the HI model as follows: we imagine starting with a large value of the lattice spacing $a$, and decreasing it slowly, and monitor the overlap integral with neighbour in the direction $\alpha$, for fixed $M$.  At some value of spacings, one of the integrals becomes non zero for the first time.  That  is the direction of  the HI model spin associated to this $M$. Values on $M$ which lie at the common boundaries of two submanifolds,  and where this direction is not unique, have zero measure.   For all smaller $a$, this overlap will increase, but by definition, all other overlaps will remain strictly zero, so long as the AOO condition is satisfied. As a simple example,  in two dimensions, the manifold $M$ is the  full circle  with $0 \leq \theta \le 2 \pi$, and  it is broken into four submanifolds  $ \big([-\pi/4, \pi/4),[\pi/4, 3\pi/4),[3\pi/4, 5 \pi/4)$ and $[5\pi/4, 7\pi/4)\big)$ which correspond to the four states of the discrete model. \\ 
 
 The graphical expansion of the partition function, for a given dimer configuration, the sites covered by dimers have a fixed $\sigma$-variable, and we sum over all orientations of arrows at sites where there is no dimer.  The full partition function has the form of the inclusion-exclusion principle.  The empty dimer configuration corresponds to all possible states of the discrete spins.  The term with one dimer is negative, and subtracts the sum of contributions where specified  one bond is doubly occupied, and others are aribitrary, summed over all possible positions of the bond. The next term corrects for the overcounting by adding the contribution where two bonds are doubly occupied, and others are arbitrary, and so on. Then the integration over submanifolds for  specified  discrete state at  each site is easily done. In the integration over the $M$-variables, restricted to specific submanifolds, sites not covered by same dimer are independent. This implies that all the $n$-point correlation functions of rotor model with AOO condition  can be expressed as linear functions  of the $n$-point correlations of ${\mathcal H}_{HI}$. We  illustrate this  below using the 1-point function. 
  
\begin{figure}[h]
\centering
\includegraphics[width=1\columnwidth]{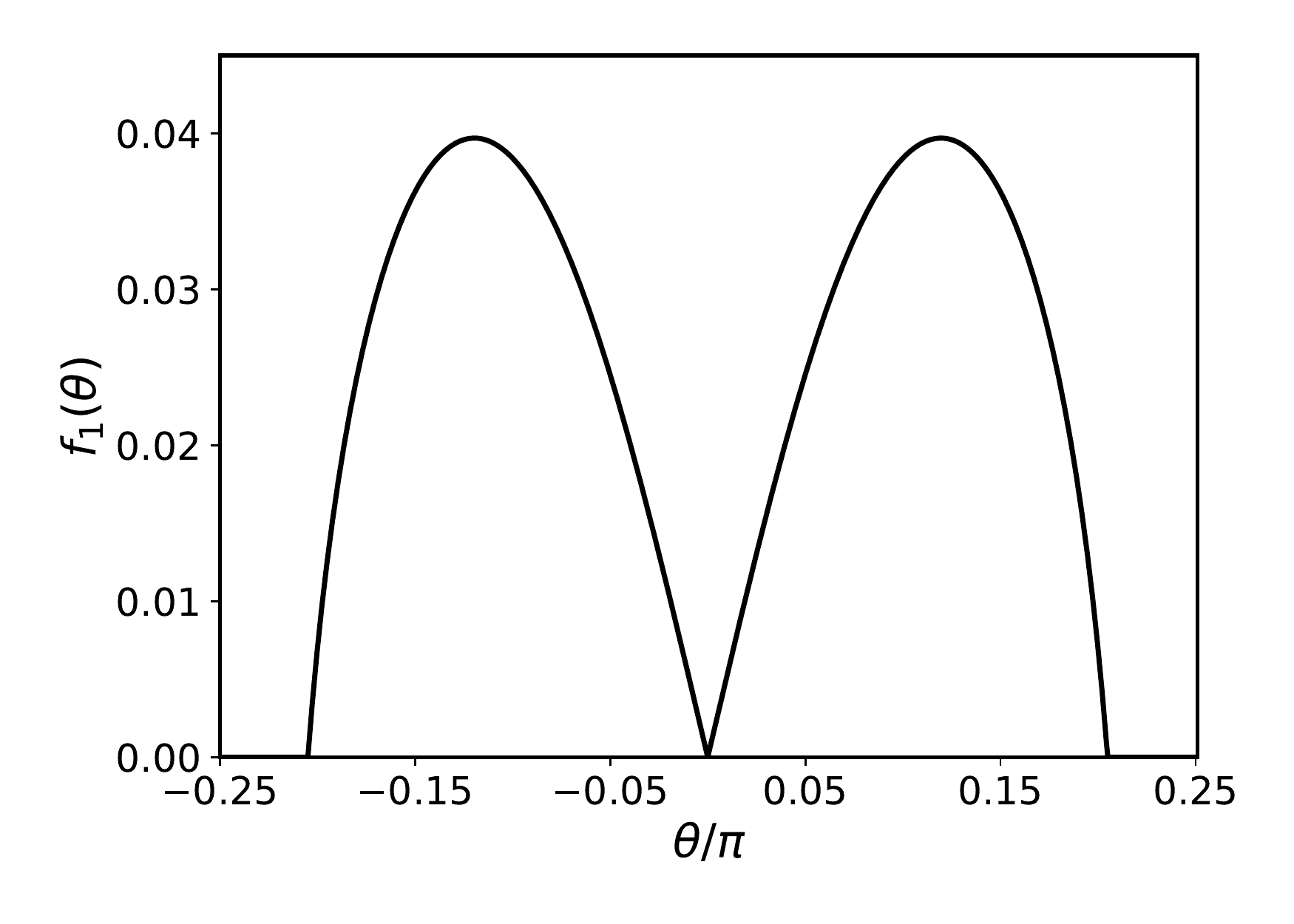}
\caption{$f_{1}(\theta)$ for a system of hard linear rods of length 2d square lattice for $a=1.6 l$ . Because of the symmetry of the square lattice  other $f_{\alpha}$ are related to $f_1$ such that  $f_{-1}(\theta)= f_1({\theta-\pi})$ and $f_{\pm 2}(\theta)= f_1({\theta \pm \pi/2})$.     }
\label{ftheta}
\end{figure} 
 Let $f_{\alpha}(M) dM$ be the conditional probability that variable $M$ of the rotor at ${\bf O}$  lies in a small volume $dM$ near the value $M$, given that it overlaps with the neighboring rotor  in the direction $\alpha$.\\
Then we have
\begin{equation}
f_{\alpha}(M) = \dfrac{\int dM({\bf \hat{e}}_{\alpha}) ~\ \eta ({\bf O},{\bf \hat{e}}_{\alpha})}{\int dM({\bf O}) \int dM({\bf \hat{e}}_{\alpha})  ~\ \eta ({\bf O},{\bf \hat{e}}_{\alpha})}, 
\label{ftheta1}
\end{equation}
In Fig. \ref{ftheta}, we have shown the plot of $f_1(\theta)$ for the linear rods pivoted at the end.  Clearly $f_{\alpha}$ is zero outside the the submanifold $\omega_{\alpha}$.\\
Let ${\rm P}(M) dM$ denote the probability that the matrix at the origin ${\bf O}$, in the thermodynamical limit, lies in the volume $dM$ centered at $M$. In the partition sum in Eq.(\ref{eq:Z}), we sum over all sites, except the origin.
If $\bar{n}(z)$ is the number density of dimers per site, then with probability weight $( 1 - 2 \bar{n}(z))$, the site is not covered by a dimer, and with weight  $\dfrac{\bar{n}(z)}{d}$, it is covered by a dimer in direction $\alpha$. Then ${\rm P}(M)$ is given by,

\begin{equation}
{\rm P}(M)=\left[ ( 1 -  2\bar{n})  + \frac{\bar{n}}{d } \sum_{\alpha}   f_\alpha( M)\right].
\label{prob1}
\end{equation}

\begin{figure}[h]
  \centering
  \includegraphics[width=1\columnwidth]{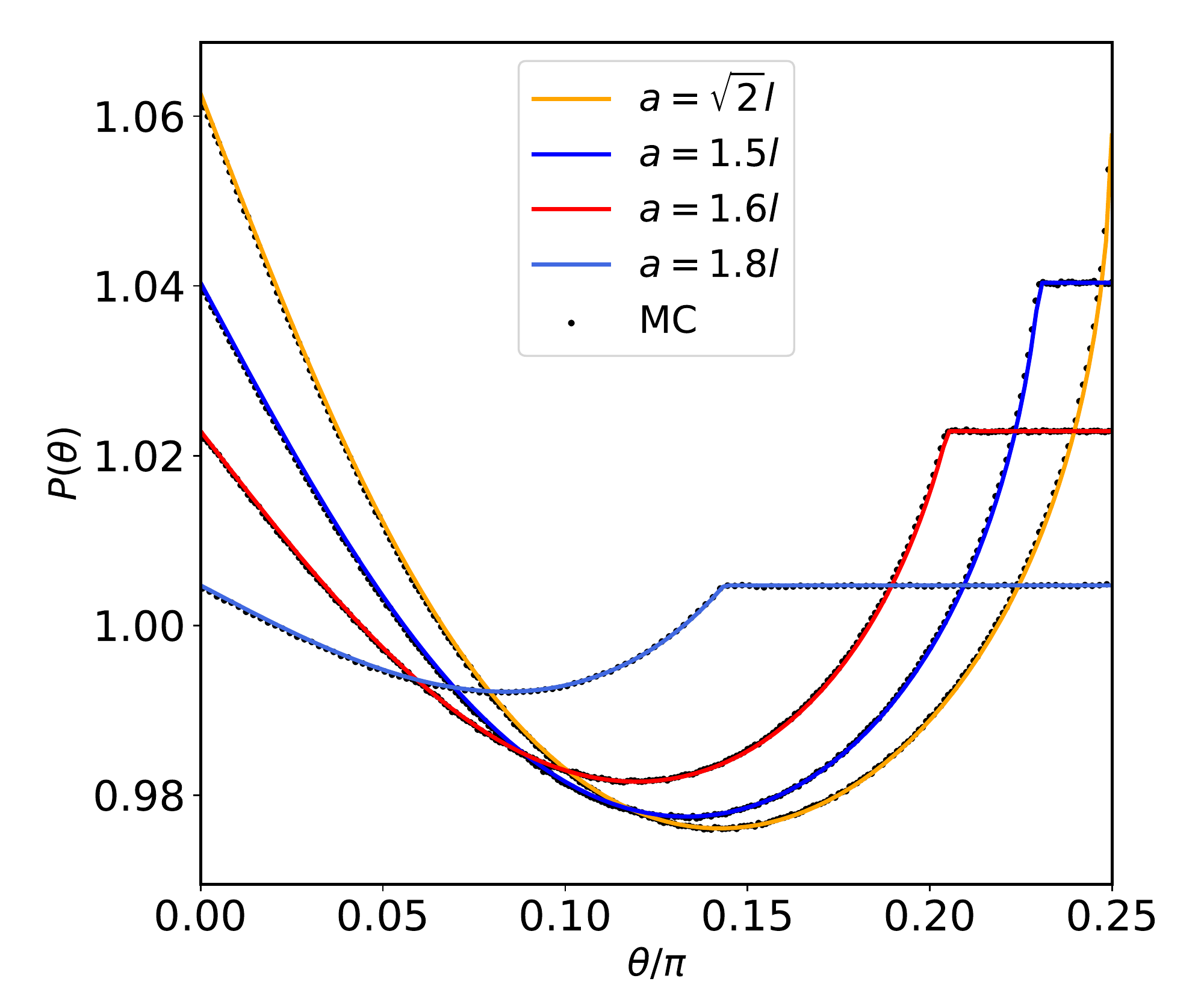}
\caption{The probability distribution of angles ${\rm P}(\theta)$ as a function of  the orientation angle $\theta$, for a system of hard rods pivoted at the one end on a square lattice (see Fig. \ref{2drods}),  for different values of lattice spacing. The function outside this range can be obtained using the symmetries of the square lattice.} 
\label{ptheta}
\end{figure}

  As a concrete example, we calculate ${\rm P}(M)$ for thin  hard  linear rods of length $\ell$  pivoted at one end at the vertices of a square lattice [Fig. \ref{2drods}]. Here $M$ is just a simple rotation angle $\theta \in [ -\pi, \pi)$. In this case, it is straight forward to calculate 
 the function $f_{\alpha}(\theta)$, for $\alpha = \pm 1, \pm 2$, numerically  for any given value of $\ell/ a$ using Eq.(\ref{ftheta1}).\\  
 In Fig. \ref{ptheta}, we have plotted the function ${\rm P}(\theta)$, as determined by Monte Carlo simulations, against the theoretical form, for different values of $\ell/a$. We do not have the functional form of $\bar{n}(z)$, as the solution to the dimer model is not known. However,  the experimental data can be fitted to the functional form in Eq.(\ref{prob1}), treating the unknown value $\bar{n}(z)$ as a single fitting parameter , for a given value of $\ell/a$.This is done in Fig. 6, and we see that  using only a single parameter, we are able to get a very good fit to the function ${\rm P}(\theta)$ for the entire  range of $\theta$, and for different values of $\ell/a$. In Appendix B and C, we verify that the distribution of orientations also agrees with that calculated using the eigenvectors of the transfer matrix for  the 1-dimensional chain, and the zig-zag chain.\\
 
\begin{figure}[H]
\centering
\includegraphics[width=1\columnwidth]{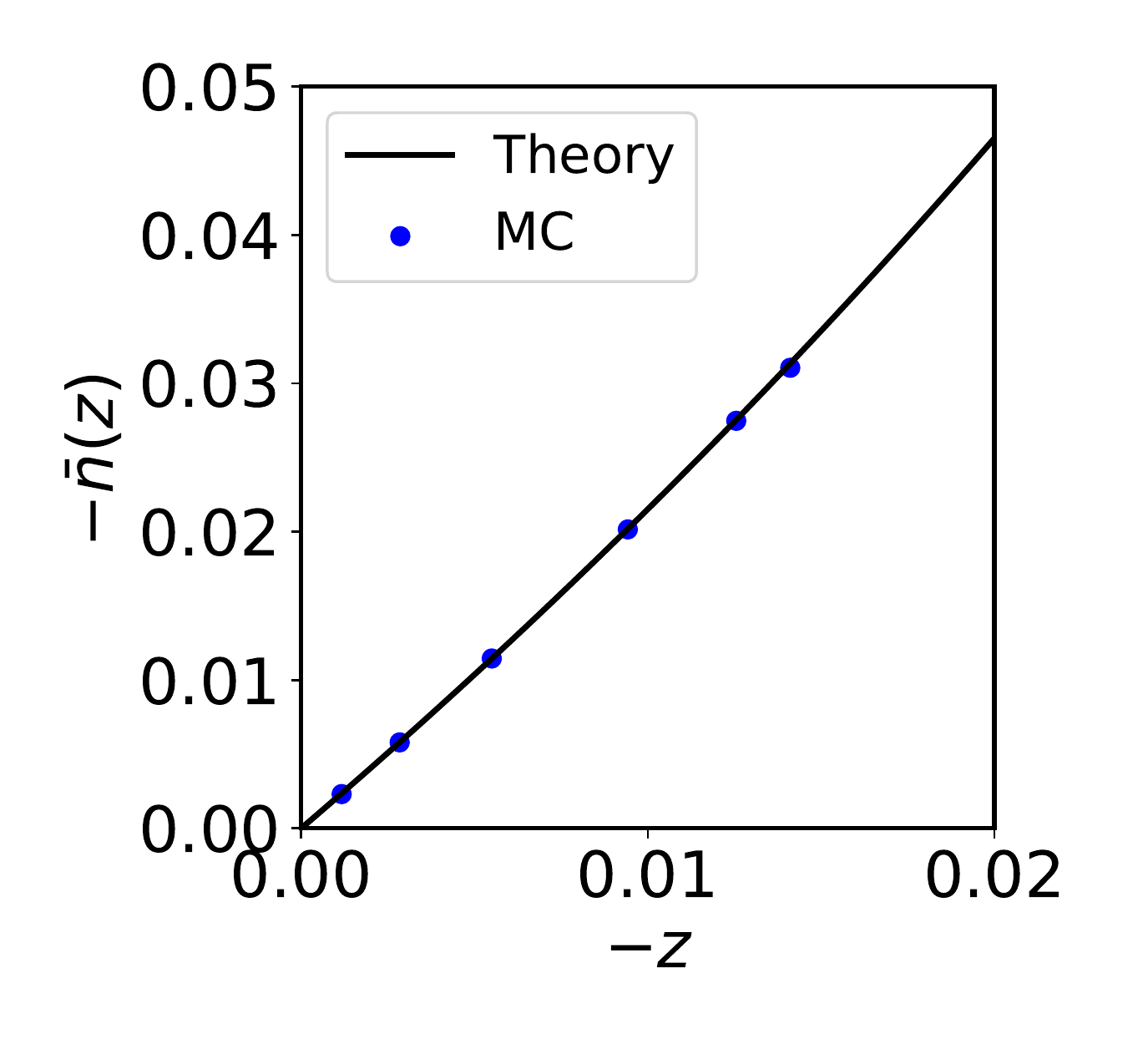}
\caption{Plot of negative of average density $\bar{n}(z)$ per site obtained from Monte Carlo simulations and theoretical value given by eq.(\ref{heap-series}). }
\label{rho}
\end{figure}

  In Fig. \ref{rho}, we have plotted the fitting value of $\bar{n}(z)$ as a function of the overlap integral $z$ for different values of $\ell/a$. As already mentioned, exact functional form of $\bar{n}(z)$ is not known.   But, for the low values of activity  in  the range where AOO holds, it is adequate to   use first few terms in the low density  Mayer series for the  dimer density as a function of the activity $z$ : 
 \begin{equation}
 \bar{n}(z) =  \sum_{n=1}^{\infty}(-1)^{n-1} a_n z^n
 \label{heap-series}
 \end{equation}

It was shown by Viennot that the Mayer coefficients $a_n$
are integers equal to the number of heaps one can form
on the simple cubic lattice using dimers \cite{vien}. For small
values of $z$ of interest here, only the first few terms of
this series are enough to determine $\bar{n}(z)$ to three digit
accuracy, which is given by
\begin{equation}
 \bar{n}(z) =   2\left[ z - 7 z^2 + 58 z^3 -  521 z^4 ....\right]
 \label{heap-series}
 \end{equation}

  At $\ell/a= 1/\sqrt{2}$, the value of activity is approximately $z = -0.014$, which is much less than the radius of convergence of the series in Eq.(\ref{heap-series}). Even though the value of $z$ is rather small, the effect on $P(\theta)$ is quite substantial as is seen in Fig. \ref{ptheta}. From the listed few terms above, we can estimate that the radius of convergence is  near   $0.1$ ( a  much more precise estimate is clearly possible, but this is adequate for our purpose  here). \\
This can be understood as follows:   We need the  AOO condition to establish the equivalence between the rotors model and the  dimer model.  The breakdown of the AOO condition is due to geometrical reasons  specific to the rotors model. The dimer model partition function  shows  no singularity  as a function of activity,  if  the  lattice spacing is decreased so that the AOO condition is no longer satisfied.\\

Of course, the rotors model is well defined, even  outside the range of lattice parameters  where the AOO condition is satisfied,  the minimum lattice spacing allowed  being set by density of closest  packing.  The mechanism leading to geometrical transitions  discussed in \cite{saryal1}  is quite robust, and holds  in all dimensions.   If we decrease $a $ below $a^*$,    we  would expect to get a sequence of geometrical transitions.  In addition,  there can be other  order-disorder type of transitions.\\
As $a$ is decreased below $a^{*} = \sqrt{2} l$,  the function z(a) is a smooth function of $a$,  but  there are configurations there one dimer overlaps with two others. These give rise to   non-zero positive correction to the  entropy $s(\ell/a)$ .  If $a = a^{*} ( 1 - \epsilon)$, this correction is of  order  $\epsilon^{3/2}$ (see Appendix D). 

\begin{figure}[H]
\centering
\includegraphics[width=0.6\columnwidth]{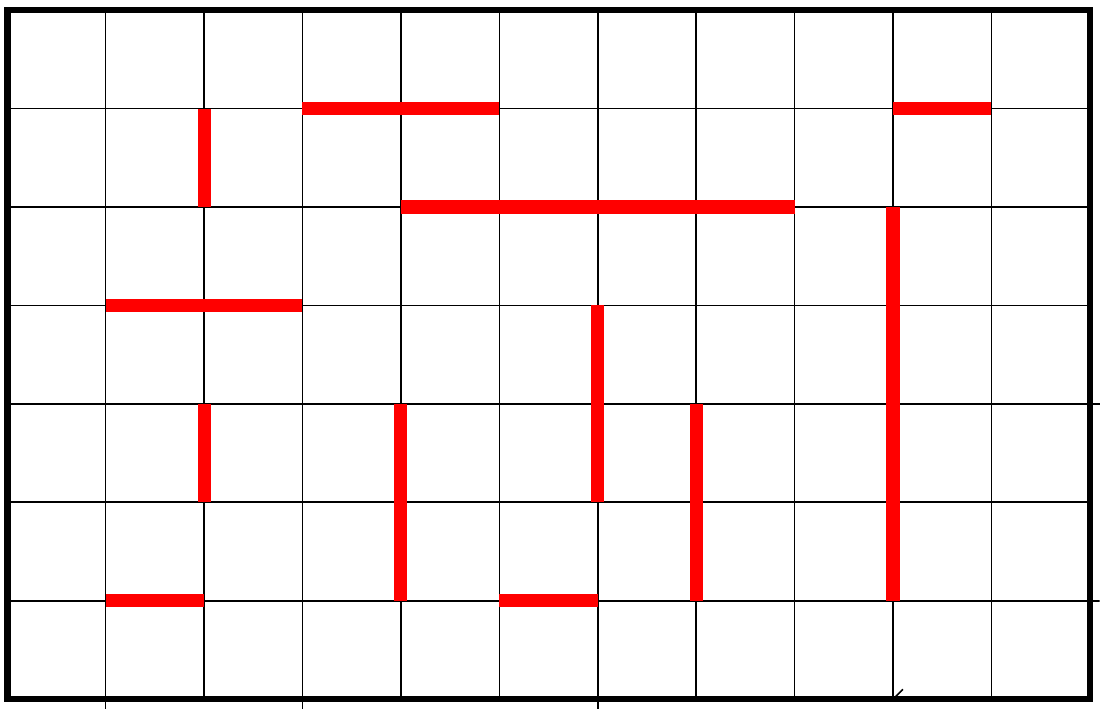}
\caption{A typical term in the graphical expansion of the partition function of rotors, when the objects are reflection symmetric, and the AOO condition is not satisfied. Here, the terms can be thoughts of as configurations in the grand partition function  of a system of polydisperse straights rods on the lattice, where the activity of a rod depends on its length.  }
\label{d2}
\end{figure}

Now, we discuss the case where the pivot point is a center of inversion of the rotor.  For example,   the two-dimensional case, when rotors are elliptical, with pivot at the center of the ellipse. In this case, AOO condition can not be satisfied on the square lattice,  but  there is a range of values of $a$, such that  an ellipse can touch two other ellipses, only if the centers are collinear.  Then, the graphical representation, terms that contribute can be thought of as configurations of non-overlapping  straight needles, whose length can be any integer.  The weight  of a needle depends on its length ( see Fig. \ref{d2}).   Thus, in this case, the partition function becomes equivalent to that of a polydisperse long rods on the same lattice. The latter problem has been studied numerically,  but remains quite intractable analytically.\\
As a final remark we want to point out that this analysis is also valid for soft rods. In case of attractive interaction among rods $z$ will be positive.\\
In summary, we have shown that there is a range of lattice spacings, where the calculation of  partition function of the  rotor model becomes exactly  equivalent to the problem of evaluating the dimer model partition function on the same lattice,   analytically continued to  negative activities of dimers. This equivalence allows us to determine the distribution of orientations for the original model. The results agree with direct Monte Carlo simulation of the  rotor model.  The results are valid for arbitrary shapes of molecules, and all dimensions. \\
 It is hoped that insights from the rather idealized  model  will  help generate  a better understanding of the actual  plastic crystals.\\

{\bf Acknowledgments}:  We thank M. Barma,   K.  Damle,  J.  Klamser, S. N. Majumdar, R. Rajesh,   T. Sadhu, and G. J. Sreejith for their comments on the paper.  DD's work is supported by a Senior Scientist fellowship from the National Academy of Sciences of India. S.S. acknowledges financial support from Council
of Scientific and Industrial Research (Grant No. 1061651988).


\renewcommand{\theequation}{A\arabic{equation}}
\setcounter{equation}{0}  

\section*{Appendix A: Calculation of $\eta(\theta,\theta^{'})$ for thin hard linear rods pivoted at one end}\label{A}
\begin{figure}[H]
\centering
\includegraphics[width=0.5\columnwidth]{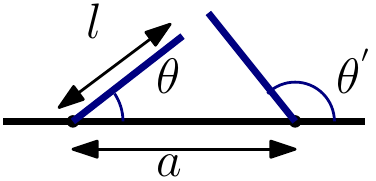}
\caption{ Two thin hard rods pivoted at adjacent lattice sites.}
\label{eta}
\end{figure}

\begin{figure}[h]
\includegraphics[width=1\columnwidth]{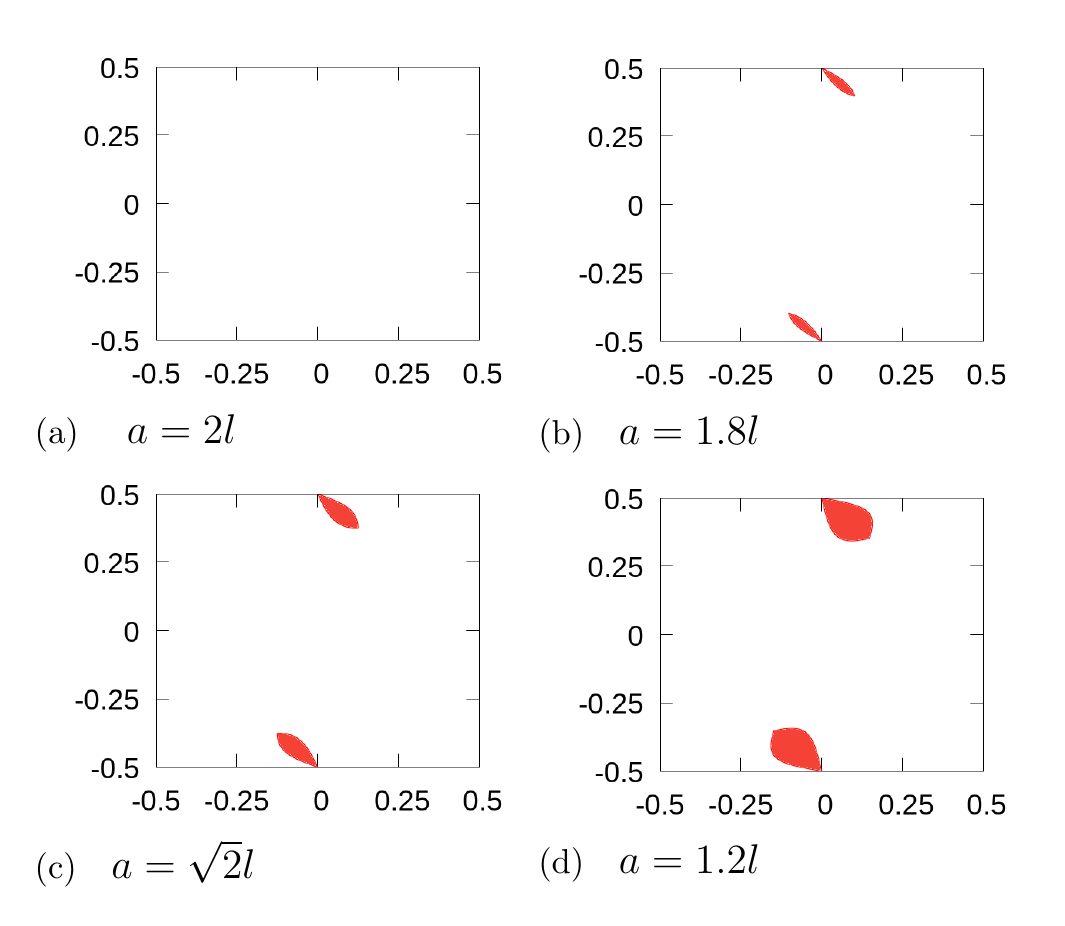}
\caption{$\eta(\theta,\theta^{'})$ in \Big($\dfrac{\theta}{2 \pi}$,$\dfrac{\theta^{'}}{2 \pi}$\Big) plane. Shaded area(red) corresponds to overlap region where $\eta$ is  $1$. In the unshaded area $\eta =0$.}
\label{transfer}
\end{figure}

Consider two thin hard rods of length $l$ placed at adjacent sites, see Fig. \ref{eta}. $\eta(\theta,\theta^{'})$ will be 1 when they overlap and 0 otherwise, see Fig. \ref{transfer}.  For notational convenience we define $\kappa=l/a$. Then, if $\kappa < 1/2$, each rod can rotate fully freely, and $\eta(\theta,\theta^{'})=0$ for all $\theta,\theta^{'}$. 
While AOO condition in 2d square lattice is valid for $1/2 \leq \kappa \leq 1/\sqrt{2}$, for 1d case, which we will discuss in next section, it is valid for $1/2 \leq \kappa \leq 1$. Therefore, we will discuss here the full range $1/2 \leq \kappa \leq 1$. \\
Because of the symmetry of the problem we have $\eta(\theta,\theta^{'}) = \eta(-\theta,-\theta^{'})$, we will only consider $\theta > 0$ to determine $\eta(\theta,\theta')$. \\

For $1/2 \le \kappa \le 1/\sqrt{2} $ , there is no overlap if $|\theta| > \cos^{-1}(\frac{1}{2\kappa})$ . If $0< \theta <  \cos^{-1}(\frac{1}{2\kappa}) $ rods will overlap only if $\theta^{'} \in [\theta^{'}_{min},\theta^{'}_{max}] $ where 
\bea
\label{max1}
\theta^{'}_{max}=\pi+\theta -\sin^{-1}\Big[\frac{\sin{\theta}}{\kappa}\Big]\\
\theta^{'}_{min}=\pi- \tan^{-1}\Big[\frac{\kappa \sin{\theta}}{1 -\kappa \cos{\theta}}\Big]
\label{min1}
\eea
For $1/\sqrt{2} \le \kappa \le 1  $, there is no overlap if $|\theta| > \sin^{-1}\Big(\kappa\Big)$. Here $\theta^{'}_{max}$ has the same expression as given in eq.(\ref{max1}) but $\theta^{'}_{min}$ has different expression for two different ranges of $\theta$. For $  \theta \in [\cos^{-1}\Big(\frac{1}{2\kappa}\Big),\sin^{-1}\Big(\kappa\Big)] $ 
\bea
\theta^{'}_{min}=\theta + \sin^{-1}\Big[\frac{\sin{\theta}}{\kappa}\Big]
\label{min2}
\eea 
For $0< \theta < \cos^{-1}(\frac{1}{2\kappa})$\,\,\, $\theta_{max}$ is given by eq.(\ref{max1}). A graphical representation of $\eta (\theta,\theta')$ is given in Fig. \ref{transfer}, for some representative values of $\ell/a$.


Given the functional form of $\eta(\theta, \theta')$, the integral in equation (\ref{eq:z}) of main text is easily done numerically, to determine $z$ as a function of $l/a$. 

\renewcommand{\theequation}{B\arabic{equation}}
\setcounter{equation}{0}  

\section*{Appendix B:  Transfer matrix calculation for the 1-dimensional chain of hard thin rods pivoted at one end}\label{B}

\begin{figure}[H]
\centering
\includegraphics[width=0.7\columnwidth]{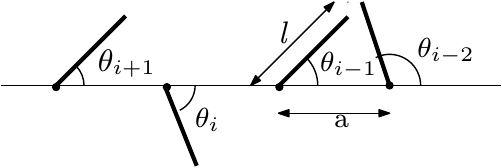}
\caption{ A system of hard rods pivoted at one end  on a one dimensional lattice.}
\label{oned}
\end{figure}

We consider  a system of thin hard rigid linear rods, each of length $\ell$. One end of the rods is fixed on regular lattice of spacing $a$, and the rods can rotate in a plane, see Fig. \ref{oned}. The orientation of each rod is specified  by the angle it makes with the x-axis.  Let the angle the $i$th rod makes be called $\theta_i$, with $ -\pi \leq \theta_i < \pi$.  

Then we partition function is given by,
\begin{equation}
\mathcal{Z}_{L} = \left[ \prod_{i} \int \frac{d\theta_i}{2 \pi} \right]  \prod_{i=1}^{L-1}\left[ 1 - \eta(\theta_{ i+1},\theta_i)\right]
\end{equation}

We will obtain the entropy per rod using the transfer matrix method, which provides a check of the algebraic method used in the main text. Also, we can obtain the probability distribution of the angles ${\rm P}(\theta)$ in terms of the eigenvector of the transfer matrix, which also agrees with the result obtained in the main text. 

AOO condition  is clearly satisfied if  $1/2 \leq \kappa \leq 1$. Let $\mathcal{Z}_{L}(\theta)$ be the restricted partition function of the system of $L$ hard rods  with left most rod having angle $\theta$. Then it satisfies the following recursion relation

\bea
\mathcal{Z}_{L+1}(\theta)=\int_{-\pi}^{  \pi} \frac{d\theta'}{2 \pi} \mathcal{T}(\theta,\theta' )\mathcal{Z}_{L}(\theta')
\label{tr1}
\eea
where  ${\mathcal T}$ is the transfer matrix.  It  is $0$ when the adjacent rods with angles ($\theta,\theta^{'}$) overlap and 1 when they don't  overlap see Fig. \ref{transfer}. Then the free energy per site in the thermodynamic limit is given by $\log(\lambda)$  where $\lambda$ is the largest eigenvalue of the transfer matrix ${\mathcal T}_{\kappa}$ , whose eigenequation is given by
\bea
\int \frac{d\theta^{'}}{2\pi} {\mathcal T}(\theta,\theta^{'}) \psi_{\lambda}^R(\theta^{'}) = \lambda\psi_{\lambda}^R(\theta)  
\label{tr1}
\eea
where $\psi_{\lambda}^R(\theta)$ is the right eigenfunction of integral kernel  ${\mathcal T}(\theta,\theta')$.  
Clearly
\begin{equation}
{\mathcal T}(\theta,\theta')= 1  -\eta(\theta,\theta').
\label{t1}
\end{equation}

Now, $\eta(\theta,\theta')$ is nonzero only if $ \cos\theta  > 0$, and $  \cos \theta' < 0$. 
Thus, we see that the AOO condition is satisfied
\begin{equation}
\eta(\theta, \theta') \eta ( \theta', \theta'') =0, {\rm ~for ~all~} \theta, \theta',\theta'',
\end{equation}

Let us define function  $f(\theta)$ given by 
\begin{equation}
f(\theta) = \int \frac{ d\theta'}{2 \pi} \eta(\theta,\theta').
\end{equation}
Clearly, $f(\theta)$ is nonzero only if, $ |\theta| <  \cos^{-1}(\frac{1}{2\kappa})$ for $1/2 \le \kappa \le \sqrt{2} $ and $ |\theta| <  \sin^{-1}(\kappa)$ for $1/\sqrt{2} \le \kappa \le 1  $   , and  it equals to $[\theta^{'}_{max}(\theta)- \theta'_{min}(\theta)]/(2 \pi)$. 
Then it is easily seen that the right eigenfunction is given by
  
\begin{equation}
 \psi_{\lambda}^R(\theta) =  1 - \frac{1}{\lambda} f(\theta),
\end{equation}
where eigenvalue $\lambda$ satisfies the equation 

\bea
\lambda^2  -\lambda + A =0
\label{eig4}
\eea

where $A$ is the fraction area of the forbidden region (red region in Fig. \ref{transfer}) , given by
\bea
A = \int  \frac{d\theta}{2\pi} \int  \frac{d\theta^{'}}{2\pi} \eta(\theta, \theta').
\eea
Thus the larger eigenvalue is given by
\begin{equation}
\lambda =\frac{1 + \sqrt{1-4 A}}{2}
\label{ev2}
\end{equation}	

Now it is easily  that for dimer model in one dimension with activity $z$, eigenvalue  of  the transfer matrix  is given by,

\begin{equation}
\lambda_{dimers} =\frac{1 + \sqrt{1+4 z}}{2}
\end{equation} 

\begin{figure}[h]
\centering
\includegraphics[width=1\columnwidth]{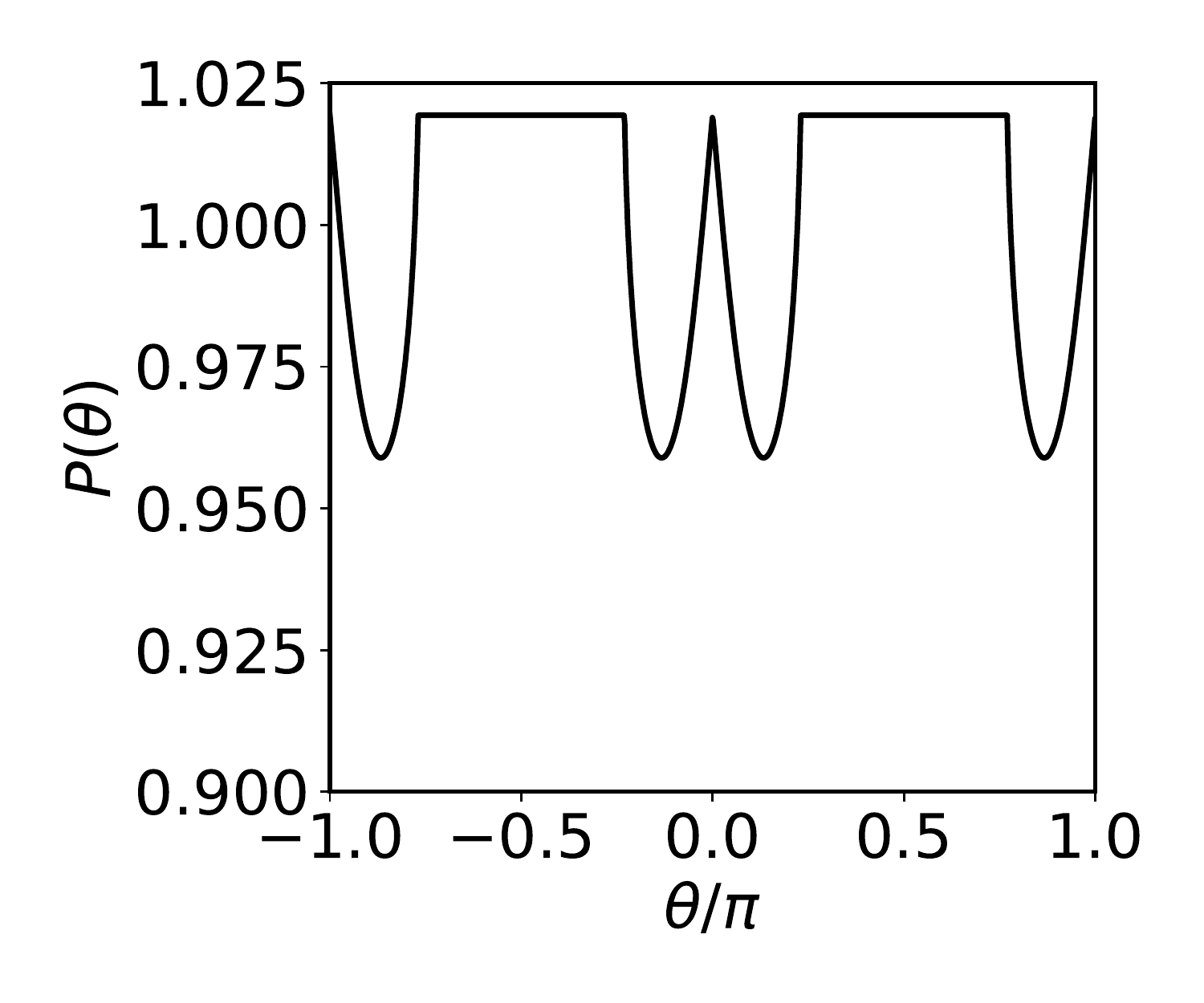}
\caption{Probability distribution of angles for 1-dimensional chain of hard thin rods pivoted at one end. $a=1.5 l$.}
\label{pthlin}
\end{figure}

Hence the model of hard rods is equivalent to the dimer model provided $z = -A$.\\
In order to obtain the probability distribution $P(\theta)$ we need both left and right eigenfunctions of the transfer matrix. Because of the symmetry of the problem left eigenfunction is simply, $\psi_{\lambda}^L(\theta)=\psi_{\lambda}^R(\theta-\pi)$. Then $P(\theta)$  is given by (see Fig. \ref{pthlin})
\bea
P(\theta)=\frac{1-\dfrac{1}{\lambda}[f(\theta)+f(\theta-\pi)]}{1+\dfrac{2}{\lambda} z}.
\eea.

\renewcommand{\theequation}{C\arabic{equation}}
\setcounter{equation}{0}  
\section*{Appendix C: Transfer matrix for the zigzag chain}\label{C}

\begin{figure}[H]
\centering
\includegraphics[width=0.9\columnwidth]{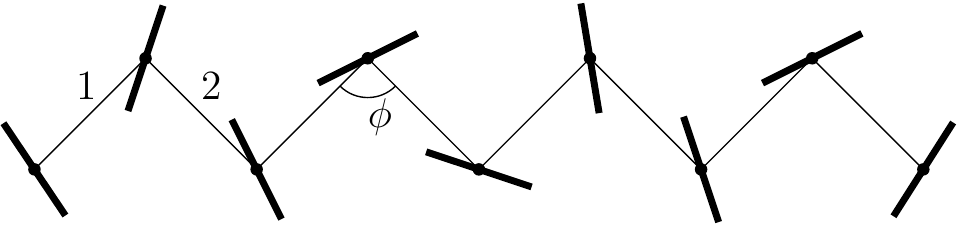}
\caption{System of hard rods pivoted at center on a 1-dimensional zigzag lattice.  Type 1 and Type 2 bonds make  ($\pi/2-\phi/2$) and (-$\pi/2+\phi/2$) angles from the x-axis respectively.}
\label{zig}
\end{figure}

We consider a slightly more complicated case of a one-dimensional zigzag lattice, where the angle between adjacent bonds alternates  between two values $ \phi$ and $ - \phi$ [ Fig. \ref{zig}]. At each site is a linear rigid thin rod of length $\ell$ pivoted to lattice site at its midpoint.
The orientation of the rod at site $i$  will be specified with its  angle $\theta_i$ it makes with one of the bond directions. We can choose the angle to lie in the range $[ -\pi/2,+\pi/2]$.
 Each rod is free to rotate about its center as long as it do not overlap with other rods.

\begin{figure}[h]
\centering
\includegraphics[scale=0.5]{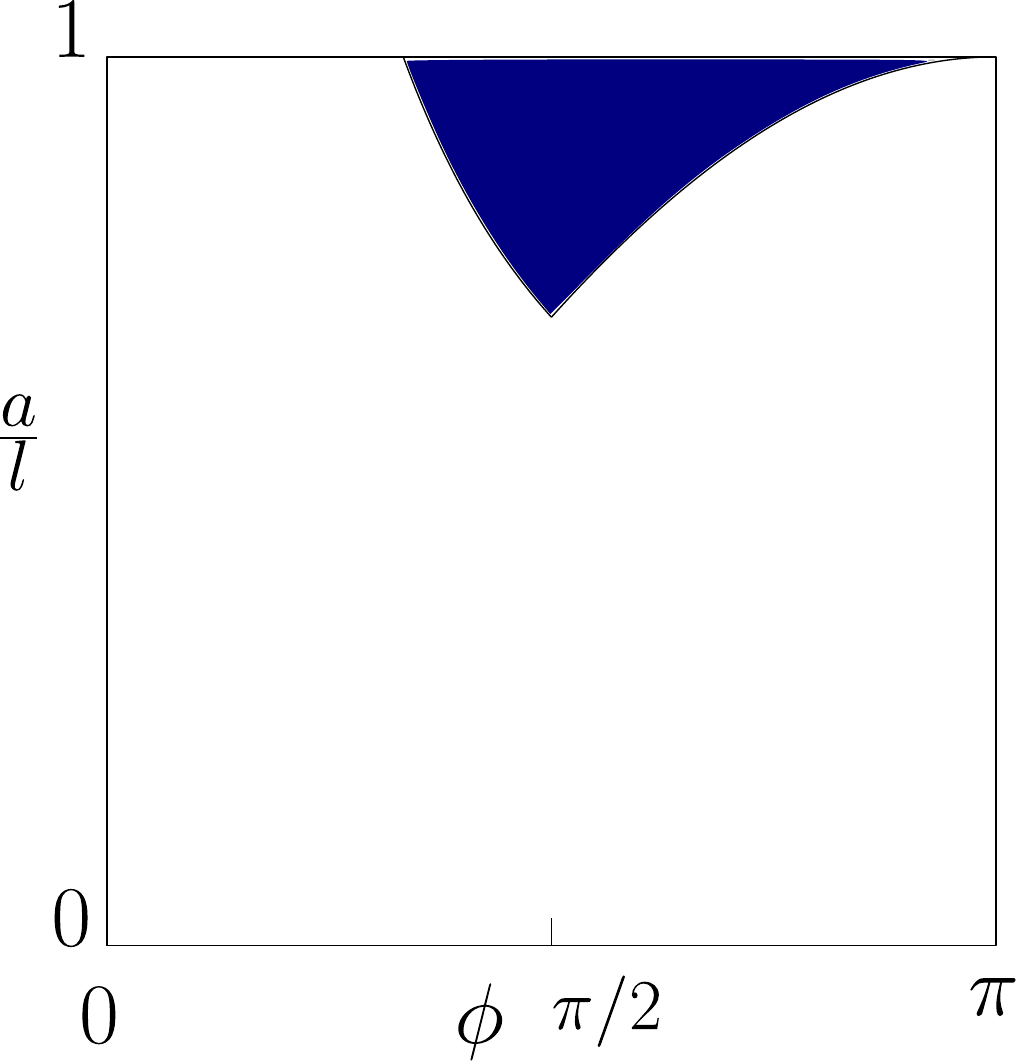}
\caption{Colored area corresponds the region where AOO condition is valid.}
\label{zigzag}
\end{figure}

 In this case, in the transfer matrix method, we get  two coupled eigenvalue equations of the form 
\bea
\label{opeg1}
\mathcal{T}_1 \ket{g}= \lambda \ket{f}\\
\mathcal{T}_2 \ket{f}= \lambda \ket{g}
\label{opeg2}
\eea 
Here ${\mathcal T}_1$ and ${\mathcal T}_2$ are the transfer matrices along the two alternating bonds. We write  $\mathcal{T}_1$ and $\mathcal{T}_2$ as 
\bea
\mathcal{T}_1= \ket{1}\bra{1}-\Delta_1 \\
\mathcal{T}_2= \ket{1}\bra{1}-\Delta_2
\eea
where $\ket{1}$ is a vector whose  value in $\theta$ representation is always one and $\Delta_{1,2}$ is the overlap operator which in $\theta$ representation takes value 1 when there is a overlap between adjacent rods and takes value zero otherwise. 
In this case, the AOO condition becomes 
\bea
\Delta_1 \Delta_2= \Delta_2 \Delta_1 =0. 
\label{cond2}
\eea
Then the following vectors are solutions of eq.(\ref{opeg1}) and eq.(\ref{opeg2})
\bea
\ket{f}=(1 -\frac{1}{\lambda} \Delta_2)\ket{1}\\
\ket{g}=(1 -\frac{1}{\lambda}\Delta_1)\ket{1}
\eea
where the  eigenvalue $\lambda$ satisfies following quadratic equation
\bea
\lambda^2-\lambda+A=0
\eea
where $A$ is , as before, given by
\bea
A &=&\braket{1|\Delta_1|1} = \braket{1|\Delta_2|1}\nonumber \\
&=& \int_{-\pi/2}^{\pi/2} \frac{d\theta}{\pi} \int_{-\pi/2}^{\pi/2} \frac{d\theta^{'}}{\pi} \bra{\theta^{'}}\Delta_{1,2}\ket{\theta}
\eea 
which is nothing but the area of the overlap region. Again we are able to obtain the same result obtained by analytic method  by using simple algebraic property of the  transfer matrix given by eq.(\ref{cond2}). In  Fig. \ref{zigzag} colored area corresponds to the region where AOO condition is satisfied in the ($a/l, \phi$) plane.  

\begin{figure}[H]
\centering
\includegraphics[width=1\columnwidth]{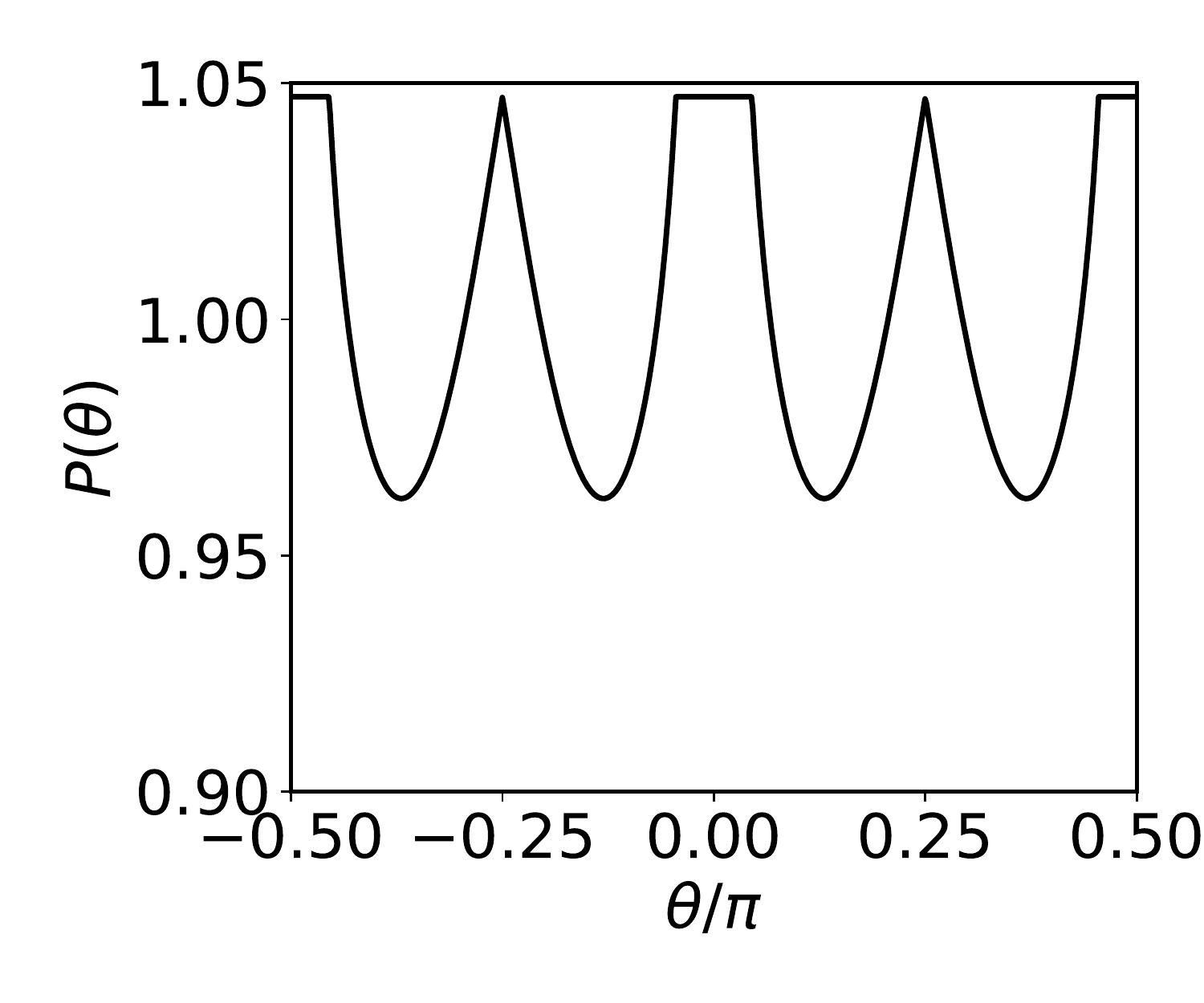}
\caption{Probability distribution for system of rods are pivoted at the center on a one-dimensional zigzag lattice. $a=0.8 l$ and $\phi=\pi/2$}
\label{pthzig}
\end{figure}
Probability distribution $P(\theta)$ can also calculated from eigenvectors, given by (see Fig. \ref{pthzig}),
\bea
P(\theta)=\frac{1-\dfrac{1}{\lambda}[\braket{\theta|\Delta_1|1}+\braket{\theta|\Delta_2|1}]}{1+\dfrac{2}{\lambda} z}.
\eea
where $z=-A$ as before.

\renewcommand{\theequation}{D\arabic{equation}}
\setcounter{equation}{0}  
\section*{ Appendix D: Qualitative behaviour of the entropy for lattice spacing just below $a^{*}$}\label{D}
\begin{figure}[H]
\centering
\includegraphics[width=0.5\columnwidth]{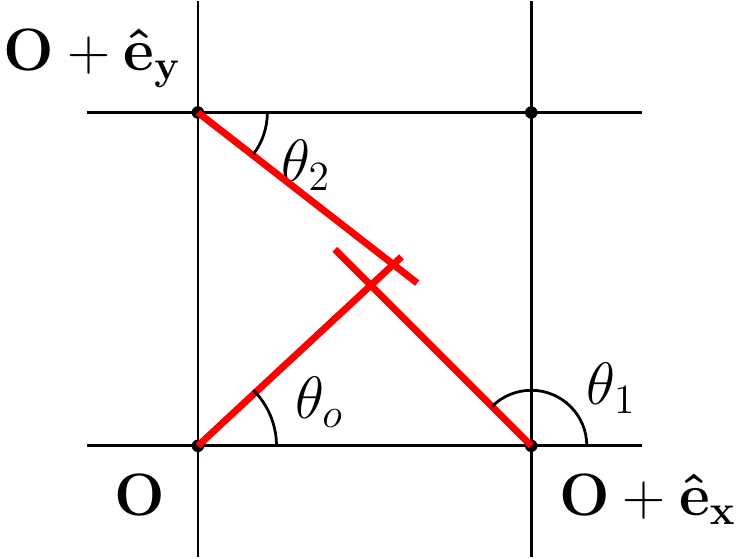}
\caption{Rotor at $\bf{O}$ overlaps simultaneously with rotors at $\bf{O}+\bf{\hat{e}_y}$ and $\bf{O}+\bf{\hat{e}_y}$.}
\label{heur1}
\end{figure}

In this section we will consider 2-dimensional model of hard rods pivoted at one end on lattice site (see Fig. 2 of main text), when lattice spacing $a=a^*(1-\epsilon) $ with $\epsilon \ll 1$. We present a heuristic argument that for this case, we have

\bea
s(l/a)&=& g(z), ~~~~~~~~a>a^* \nonumber \\
&=& g(z)+ k \epsilon^{3/2} + \textrm{higher order terms in}~\epsilon, a<a^* \nonumber \\
\label{aoov}
\eea

where $z$ is the dimer activity corresponding to lattice spacing $a$ given by eq.(6) of main text.\\
If AOO condition is violated, there are configuration where one rotor overlaps with two neighbours. A typical case is shown in Fig. \ref{heur1}

The contribution of such configurations to the partition function is of the form, $$\int \frac{d\theta_o}{2\pi}\,\int \frac{d\theta_1}{2\pi}\,\int \frac{d\theta_2}{2\pi} \eta(\theta_o,\theta_1)\,\eta(\theta_o,\theta_2). $$ 
It is clear from the geometry of the figure that non-zero contribution to the integral happens only when, ($\theta_o-\pi/4$), ($\theta_1-3\pi/4$) and ($\theta_2+\pi/4$) must be of order $\epsilon^{1/2}$ and thus integral is of order $\epsilon^{3/2}$.This integral makes positive contribution to the partition function. Hence we get the result given in eq.(\ref{aoov}).

\end{document}